\documentclass[amssymb,nobibnotes,superscriptaddress,longbibliography,twocolumn]{revtex4-1}

\usepackage{enumerate,amsmath}
\usepackage{graphicx}
\usepackage{color}

\newcommand{\mb}[1]{\mbox{\boldmath $#1$}}

\begin{document}

\title{Quantum reservoir computing: a reservoir approach toward quantum machine learning
on near-term quantum devices}%

\author{Keisuke Fujii}
\email{fujii@qc.ee.es.osaka-u.ac.jp}
\address{Graduate School of Engineering Science, Osaka University, 1-3 Machikaneyama, Toyonaka, Osaka 560-8531, Japan.}

\author{Kohei Nakajima}
\email{k\_nakajima@mech.t.u-tokyo.ac.jp}
\address{Graduate School of Information Science and Technology, The University of Tokyo, Bunkyo-ku, 113-8656 Tokyo, Japan}

\date{\today}
\begin{abstract}
Quantum systems have an exponentially large degree of 
freedom in the number of particles and hence provide a rich 
dynamics that could not be simulated on conventional computers.
Quantum reservoir computing is an approach to use such a complex and rich dynamics 
on the quantum systems as it is for temporal machine learning. 
In this chapter, we explain quantum reservoir computing and related approaches,
quantum extreme learning machine and quantum circuit learning,
starting from a pedagogical introduction to quantum mechanics and 
machine learning. 
All these quantum machine learning approaches are experimentally feasible and 
effective on the state-of-the-art quantum devices.
\end{abstract}


\maketitle
\section{Introduction}
Over the past several decades, we have enjoyed 
exponential growth of computational power, namely, 
Moore's law. Nowadays even smart phone or tablet PC is much more powerful 
than super computers in 1980s.
Even though, people are still seeking  more computational power,
especially for artificial intelligence (machine learning), 
chemical and material simulations, 
and forecasting complex phenomena like economics, weather and climate.
In addition to improving computational power of conventional computers,
i.e., more Moore's law,
a new generation of computing paradigm 
has been started to be investigated to go beyond Moore's law.
Among them, natural computing seeks to exploit natural physical or biological 
systems as computational resource.
Quantum reservoir computing is an intersection of 
two different paradigms of natural computing, namely,
quantum computing and reservoir computing. 

Regarding quantum computing, 
the recent rapid experimental progress in controlling complex quantum systems
motivates us to use quantum mechanical law as a new principle of information processing,
namely, quantum information processing~\cite{NielsenChuang,FujiiText}.
For example,  certain mathematical problems, such as integer factorisation, which are believed to be intractable on a classical computer, are known to be efficiently solvable by a sophisticatedly synthesized quantum algorithm~\cite{ShorFact}.
Therefore, considerable experimental effort has been devoted to realising full-fledged universal quantum computers~\cite{UCSB1,UCSB2}.
In the near feature, 
quantum computers of size $>50$ qubits with 
fidelity $>99\%$ for each elementary gate would appear 
to achieve quantum computational supreamcy 
beating simulation on the-state-of-the-art 
classical supercomputers~\cite{Preskill,Boixo}.
While this does not directly mean that a quantum computer outperforms 
classical computers for a useful task like machine learning,
now applications of such a near-term quantum device
for useful tasks including machine leanring has been widely explored.
On the other hand, quantum simulators are thought to be much easier to implement than a full-fledged universal quantum computer.
In this regard, existing quantum simulators have already shed new light on the physics of complex many-body quantum systems~\cite{QuantumSimulator1,QuantumSimulator2,QuantumSimulator3}, and a restricted class of quantum dynamics, known as adiabatic dynamics, has also been applied to combinatorial optimisation problems~\cite{Nishimori,Farhi,Dwave,Dwave2}.
However, complex real-time quantum dynamics, which is one of the most difficult tasks for classical computers to simulate
~\cite{FujiiMorimae,Fujiietal,FujiiTamate} 
and has great potential to perform nontrivial information processing, is now waiting to be harnessed as a resource for more general purpose information processing.

Physical reservoir computing, 
which is the main subject throughout this book, is another paradigm for 
exploiting complex physical systems for information processing.
In this framework, the low-dimensional input is projected to a high-dimensional dynamical system, which is typically referred to as a {\it reservoir}, generating transient dynamics that facilitates the separation of input states \cite{Transient}.
If the dynamics of the reservoir involve both adequate memory and nonlinearity \cite{Capacity}, emulating nonlinear dynamical systems only requires adding a linear and static readout from the high-dimensional state space of the reservoir.
A number of different implementations of reservoirs have been proposed, such as abstract dynamical systems for echo state networks (ESNs) \cite{Jaeger0} or models of neurons for liquid state machines \cite{Maass0}.
The implementations are not limited to programs running on the PC but also include physical systems, such as the surface of water in a laminar state \cite{Bucket}, analogue circuits and optoelectronic systems \cite{Laser0,Laser1,Laser1b,Laser2,Laser3,Laser4}, and neuromorphic chips \cite{Neuromorphic0}.
Recently, it has been reported that the mechanical bodies of soft and compliant robots have also been successfully used as a reservoir \cite{Helmut0,Kohei0,Kohei1,Kohei2,Kohei3,Ken1}.
In contrast to the refinements required by learning algorithms, such as in deep learning \cite{DL}, the approach followed by reservoir computing, especially when applied to real systems, is to find an appropriate form of physics that exhibits rich dynamics, thereby allowing us to outsource a part of the computation.

Quantum reservoir computing (QRC) was born in the marriage of quantum computing and 
physical reservoir computing above to harness complex quantum dynamics as a reservoir 
for real-time machine learning tasks~\cite{QRC}.
Since the idea of QRC has been proposed in Ref.~\cite{QRC},
its proof-of-principle experimental demonstration for non temporal tasks~\cite{QRC_ex} and performance analysis and improvement~\cite{QRC_multiplex,QRC_OTOC,QRC_Higher_Order}
has been explored.
The QRC approach to quantum tasks such as quantum tomography and quantum state preparation has been recently garnering attention \cite{QRC_sig,QRC_sig1,QRC_sig2}.
In this book chapter, 
we will provide a broad picture of QRC and related approaches
starting from a pedagogical introduction to quantum mechanics and 
machine learning.

The rest of this paper is organized as follows.
In Sec~\ref{sec:quantum_mechanics},
we will provide a pedagogical introduction to quantum mechanics
for those who are not familiar to it and fix our notation.
In Sec~\ref{sec:machine_learning}, we will briefly mention to 
several machine learning techniques like,
linear and nonlinear regressions, temporal machine learning tasks
and reservoir computing.
In Sec~\ref{sec:quantum_machine_learning},
we will explain QRC and related approaches,
quantum extreme learning machine~\cite{QRC_ex} and quantum circuit learning~\cite{QCL}.
The former is a framework to use quantum reservoir for non temporal tasks,
that is, the input is fed into a quantum system, and generalization or classification 
tasks are performed by a linear regression on a quantum enhanced feature space.
In the latter, the parameters of the quantum system is further fine-tuned 
via the gradient descent by measuring an analytically obtained gradient,
just like the back propagation for feedforward neural networks.
Regarding QRC, we will also see chaotic time series predictions as demonstrations.
Sec.~\ref{sec:conclusion} is devoted to conclusion and discussion.

\section{Pedagogical introduction to quantum mechanics} 
\label{sec:quantum_mechanics}
In this section, we would like to provide a 
pedagogical introduction to how quantum mechanical systems work
for those who are not familiar to quantum mechanics.
If you already familiar to quantum mechanics and its notations,
please skip to Sec.~\ref{sec:machine_learning}.

\subsection{Quantum state}
A state of a quantum system is described by a {\it state vector},
\begin{eqnarray}
|\psi \rangle = 
\left(
\begin{array}{c}
c_1 
\\
\vdots
\\
c_d
\end{array}
\right)
\end{eqnarray}
on a complex $d$-dimensional system $\mathbb{C}^d$,
where the symbol $| \cdot \rangle$ is called {\it ket}
and indicates a complex column vector.
Similarly, $\langle \cdot |$ is called {\it bra}
and indicates a complex row vector, and they are 
related complex conjugate,
\begin{eqnarray}
\langle \psi | = |\psi \rangle ^{\dag}  = 
\left(
\begin{array}{ccc}
c^*_1 & \cdots & c^*_d
\end{array}
\right).
\end{eqnarray}
With this notation, 
we can writte an inner product of two quantum state $|\psi \rangle$ and $|\phi \rangle$
by $\langle \psi | \phi \rangle$.
Let us define an orthogonal basis 
\begin{eqnarray}
|1\rangle 
= 
\left(
\begin{array}{c}
1
\\
0
\\
\vdots
\\
\vdots
\\
0
\end{array}
\right),...
\;\;\;
|k\rangle 
= 
\left(
\begin{array}{c}
0
\\
\vdots
\\
1
\\
0
\\
\vdots
\end{array}
\right),...
\;\;\;
|d\rangle 
= 
\left(
\begin{array}{c}
0
\\
\vdots
\\
\vdots
\\
\vdots
\\

d
\end{array}
\right),
\end{eqnarray}
a quantum state in the $d$-dimensional system 
can be described simply by
\begin{eqnarray}
|\psi \rangle = \sum _{i=1}^{d} c_i | i\rangle .
\end{eqnarray}
The state is said to be a {\it superposition state} of $|i\rangle$.
The coefficients $\{c_i\}$ are complex,
and called {\it complex probability amplitudes}.
If we measure the system in the basis $\{ |i \rangle \}$,
we obtain the measurement outcome $i$ with 
a probability 
\begin{eqnarray}
p_i = | \langle i | \psi \rangle |^2 =  | c_i | ^2 ,
\end{eqnarray}
and hence the complex probability amplitudes have 
to be normalized as follows
\begin{eqnarray}
|\langle \psi | \psi \rangle |^2 = \sum _{i=1}^d |c_i|^2 =1.
\end{eqnarray}
In other words,
a quantum state is represented as a normalized vector 
on a complex vector space.

Suppose the measurement outcome $i$ 
corresponds to a certain physical value $a_i$,
like energy, magnetization and so on,
then the expectation value of the physical valuable is given by
\begin{eqnarray}
\sum _i a_i p_i = \langle \psi | A |\psi \rangle \equiv \langle A \rangle,
\end{eqnarray}
where we define an hermitian operator
\begin{eqnarray}
A = \sum _i a_i |i\rangle \langle i|,
\end{eqnarray}
which is called {\it observable},
and has the information of the measurement basis and physical valuable.

The state vector in quantum mechanics
is similar to a probability distribution, but 
essentially different form it, since 
it is much more primitive; 
it can take complex value and 
is more like
a square root of a probability.
The unique features of the quantum systems 
come from this property.

\subsection{Time evolution}
The time evolution of a quantum system is determined by 
a Hamiltonian $H$,
which is a hermitian operator acting on the system.
Let us denote a quantum state at time $t=0$ by $|\psi (0)\rangle$.
The equation of motion for quantum mechanics,
so-called Schr{\"o}dinger equation, is 
given by
\begin{eqnarray}
i \frac{\partial}{\partial t} |\psi (t)\rangle = H |\psi (t)\rangle .
\end{eqnarray}
This equation can be formally solved by 
\begin{eqnarray}
|\psi (t) \rangle = e^{-i H t } |\psi (0)\rangle.
\end{eqnarray}
Therefore the time evolution is given by 
an operator $e^{-i H t}$,
which is a unitary operator and hence the norm of the state vector is preserved,
meaning the probability conservation.
In general, the Hamiltonian can be time dependent.
Regarding the time evolution,
if you are not interested in the continuous time evolution,
but in just its input and output relation,
then the time evolution is nothing but a unitary operator $U$
\begin{eqnarray}
|\psi_{\rm out} \rangle  = U |\psi _{\rm in} \rangle .
\end{eqnarray}
In quantum computing, the time evolution $U$ is sometimes called 
{\it quantum gate}.

\subsection{Qubits}
The smallest nontrivial quantum system is a two-dimensional quantum system $\mathbb{C}^2$, which is 
called {\it quantum bit} or {\it qubit}:
\begin{eqnarray}
\alpha |0\rangle + \beta |1\rangle , \;\;\; (|\alpha|^2 + |\beta |^2 = 1).
\end{eqnarray}
Suppose we have $n$ qubits. 
The $n$-qubit system is defined by a tensor product space $(\mathbb{C}^2)^{\otimes n}$ of each two-dimensional system
as follows.
A basis of the system 
is defined by a direct product of a binary state $|x_k\rangle$ with $x_k \in \{ 0,1\}$,
\begin{eqnarray}
|x_1 \rangle \otimes | x_2 \rangle \otimes \cdots \otimes |x_n \rangle,
\end{eqnarray}
which is simply denoted by 
\begin{eqnarray}
|x_1  x_2  \cdots x_n \rangle.
\end{eqnarray}
Then a state of the $n$-qubit system can be described as
\begin{eqnarray}
|\psi \rangle = \sum _{x_1 ,x_2 ,...,x_n} \alpha _{x_1 ,x_2, ...,x_n} |x_1  x_2  \cdots x_n \rangle.
\end{eqnarray}
The dimension of the $n$-qubit system is $2^n$,
and hence the tensor product space is nothing but 
a $2^n$-dimensional complex vector space $\mathbb{C}^{2^n}$.
The dimension of the $n$-qubit system increases exponentially in the number $n$ of the qubits.

\subsection{Density operator}
Next, I would like to introduce operator formalism of the above quantum mechanics.
This describes an exactly the same thing but sometimes the operator formalism would be convenient.
Let us consider an operator $\rho$ constructed from the state vector $|\psi \rangle$:
\begin{eqnarray}
\rho = |\psi \rangle \langle \psi |.
\end{eqnarray}
If you chose the basis of the system $\{ |i \rangle \}$ for the matrix representation,
then the diagonal elements of $\rho$ corresponds 
the probability distribution $p_i = |c_i |^2$ when the system is measured in the basis $\{ |i \rangle\}$.
Therefore the operator $\rho$ is called a density operator.
The probability distribution can also be given in terms of $\rho$ by 
\begin{eqnarray}
p_i = {\rm Tr}[ |i \rangle \langle i | \rho ], 
\end{eqnarray}
where ${\rm Tr}$ is the matrix trace.
An expectation value of an observable $A$ is given by 
\begin{eqnarray}
\langle A \rangle = {\rm Tr}[A\rho].
\end{eqnarray}
The density operator can handle a more general situation where
a quantum state is sampled form
a set of quantum states $\{ |\psi _k \rangle\}$ 
with a probability distribution $\{ q_k \}$.
In this case, if we measure the system in the basis $\{ | i\rangle \langle i|\}$,
the probability to obtain the measurement outcome $i$ is given by 
\begin{eqnarray}
p_i = \sum _k q_k {\rm Tr}[|i \rangle \langle i | \rho _k],
\end{eqnarray}
where $\rho _k = | \psi _k \rangle \langle \psi _k |$.
By using linearity of the trace function, this reads
\begin{eqnarray}
p_i = {\rm Tr}[|i \rangle \langle i | \sum _k q_k \rho _k].
\end{eqnarray}
Now we interpret that the density 
operator is given by
\begin{eqnarray}
\rho = \sum _k q_k |\psi _k \rangle \langle \psi _k | .
\end{eqnarray}
In this way, a density operator can represent
classical mixture of quantum states by a convex mixture of density operators,
which is convenient in many cases.
In general, a positive and hermitian operator $\rho$ 
being subject to ${\rm Tr} [\rho ] = 1$ can be a density operator,
since it can be interpreted as a convex mixture of quantum states 
via spectral decomposition:
\begin{eqnarray}
\rho = \sum \lambda _i | \lambda _i \rangle \langle \lambda _i | ,
\end{eqnarray}
where $\{ |\lambda _i \rangle \}$ and $\{ \lambda _i \}$ are 
the eigenstates and eigenvectors respectively.
Because of ${\rm Tr}[\rho ] =1$, we have $\sum _i \lambda _i =1$.

From its definition, the time evolution of $\rho$ can be given by 
\begin{eqnarray}
\rho (t) = e^{-i H t} \rho (0) e^{iHt}
\end{eqnarray}
or
\begin{eqnarray}
\rho _{\rm out} = U \rho _{\rm in} U^{\dag}.
\end{eqnarray}
Moreover, 
we can define
more general operations for the density operators.
For example, 
if we apply unitary operators $U$ and $V$ with probabilities $p$ and $(1-p)$,
respectively, 
then we have
\begin{eqnarray}
\rho _{\rm out} = p U \rho U ^{\dag} + (1-p) V \rho V ^{\dag}.
\end{eqnarray}
As another example, if we perform the measurement of $\rho$ in the basis $\{ |i \rangle\}$,
and we forget about the measurement outcome,
then the state is now given by a density operator
\begin{eqnarray}
\sum _i {\rm Tr}[|i \rangle \langle i | \rho ] |i \rangle \langle i | 
= \sum _i |i \rangle \langle i | \rho |i \rangle \langle i |.
\end{eqnarray}
Therefore if we define a map from a density operator to another,
which we call {\it superoperator},
\begin{eqnarray}
\mathcal{M} (\cdots) = \sum _i |i \rangle \langle i | (\cdots) |i \rangle \langle i |,
\end{eqnarray}
the above non-selective measurement (forgetting about the measurement outcomes)
is simply written by
\begin{eqnarray}
\mathcal{M} (\rho ).
\end{eqnarray}
In general, any physically allowed quantum operation $\mathcal{K}$ that maps a density operator to another
can be represented in terms of a set of operators $\{K_i \}$
being subject to $K_i ^{\dag} K_i = I$ with an identity operator $I$:
\begin{eqnarray} 
\mathcal{K}(\rho) = \sum _i K_i \rho K^{\dag}_i.
\end{eqnarray}
The operators $\{K_i\}$ are called {\it Kraus operators}.

\subsection{Vector representation of density operators}
\label{subsec:vector_density}
Finally, we would like to introduce a vector representation of the above operator formalism.
The operators themselves satisfy axioms of the linear space.
Moreover, we can also define an inner product for two operators,
so-called {\it Hilbert-Schmidt inner product}, by
\begin{eqnarray}
{\rm Tr} [A^{\dag} B ] .
\end{eqnarray}
The operators on the $n$-qubit system can be 
spanned by the tensor product of Pauli operators $\{ I, X, Y,Z\}^{\otimes n}$,
\begin{eqnarray}
P(\mb{i}) =\bigotimes_{k=1}^{n} \sigma_{i_{2k-1}i_{2k}}.
\end{eqnarray}
where 
$\sigma_{ij}$ is the Pauli operators:
\begin{eqnarray*}
I=\sigma_{00}=
\left(\begin{array}{cc}
1 & 0
\\
0 & 1
\end{array}
\right)
, X= \sigma _{10}=
\left(\begin{array}{cc}
0 & 1
\\
1 & 0
\end{array}
\right),
\end{eqnarray*}
\begin{eqnarray}
Z=\sigma _{01}=
\left(\begin{array}{cc}
1 & 0
\\
0 & -1
\end{array}
\right), Y=\sigma _{11}=
\left(\begin{array}{cc}
0 & -i
\\
i & 0
\end{array}
\right).
\end{eqnarray}
Since the Pauli operators 
constitute a complete basis on the operator space,
any operator $A$ can be decomposed into a linear combination of $P(\mb{i})$,
\begin{eqnarray}
A = \sum _ {\mb{i}} a_{\mb{i}} P(\mb{i}).
\end{eqnarray}
The coefficient $a_{\mb{i}}$ can be calculated 
by using the Hilbert-Schmidt inner product as follows:
\begin{eqnarray} 
a_{\mb{i}} = {\rm Tr}[P(\mb{i})A]/2^n,
\end{eqnarray}
by virtue of the orthogonality
\begin{eqnarray}
{\rm Tr} [P(\mb{i})P(\mb{j})] /2^n = \delta _{\mb{i},\mb{j}}.
\end{eqnarray}
The number of the $n$-qubit Pauli operators $\{ P (\mb{i})\}$
is $4^n$, and hence 
a density operator $\rho$ of the $n$-qubit system can be represented 
as a $4^n$-dimensional vector
\begin{eqnarray}
\mb{r} = 
\left(
\begin{array}{c}
r_{00...0}
\\
\vdots
\\
r_{11...1}
\end{array}
\right),
\end{eqnarray}
where $r_{00...0}=1/2^n$ because of ${\rm Tr}[\rho]=1$.
Moreover, because $P(\mb{i})$ is hermitian, 
$\mb{r}$ is a real vector.
The superoperator $\mathcal{K}$ is a linear map 
for the operator, and hence can be 
represented as a matrix acting on the vector $\mb{r}$:
\begin{eqnarray}
\rho' = \mathcal{K}(\rho) \Leftrightarrow \mb{r}' = K \mb{r},
\end{eqnarray}
where the matrix element is given by 
\begin{eqnarray}
K_{\mb{i}\mb{j}} = {\rm Tr}[ P(\mb{i}) \mathcal{K} \left(P(\mb{j}) \right)]/2^n .
\end{eqnarray}
In this way,
a density operator $\rho$ and a quantum operation $\mathcal{K}$
on it can be represented by a vector $\mb{r}$ and a matrix $K$, respectively.

\section{Machine learning and reservoir approach}
In this section, we briefly introduce machine learning 
and reservoir approaches.

\label{sec:machine_learning}
\subsection{Linear and nonlinear regression}
A supervised machine learning is a task to construct 
a model $f(x)$ from a given set of teacher data $\{ x^{(j)},y^{(j)} \}$
and to predict the output of an unknown input $x$.
Suppose $x$ is a $d$-dimensional data, and $f(x)$ is one dimensional, 
for simplicity.
The simplest model is {\it linear regression},
which models $f(x)$ as a linear function with respect to the input:
\begin{eqnarray}
f(x) = \sum_{i=1}^{d} w_i x_i + w_0.
\end{eqnarray}
The weights $\{w_i\}$ and bias $w_0$ are 
chosen such that an error between
$f(x)$ and the output of the teacher data,
i.e. {\it loss}, becomes minimum.
If we employ a quadratic loss function
for given teacher data $ \{ \{x^{(j)}_i\} , y^{(j)}\}$,
the problem we have to solve is as follows:
\begin{eqnarray}
 \min _{\{w_i\}} \sum _j (\sum_{i=0}^{d} w_i x^{(j)}_i  - y^{(j)})^2,
\end{eqnarray}
where we introduced a constant node $x_0 =1$.
This corresponds to solving a superimposing equations:
\begin{eqnarray}
\mathbf{y}  = \mathbf{X} \mathbf{w},
\end{eqnarray}
where $\mathbf{y} _j = y^{(j)}$, 
$\mathbf{X}_{ji} = x^{(j)}_i$, and
$\mathbf{w}_i =  w_i$.
This can be solved by using the Moore-Penrose pseudo inverse $\mathbf{X}^{+}$,
which can be defined from the singular value decomposition of 
$\mathbf{X} = U D V^{T}$ to be 
\begin{eqnarray}
\mathbf{X}^{+} = V D U^{T}.
\end{eqnarray}

Unfortunately, the linear regression results in a poor performance 
in complicated machine learning tasks, and any kind of nonlinearity 
is essentially required in the model.
A neural network is a way to introduce nonlinearity to the model,
which is inspired by the human brain.
In the neural network,
the $d$-dimensional input data $x$ is fed into
$N$-dimensional hidden nodes with an $N\times d$ input matrix $W^{\rm in}$:
\begin{eqnarray} 
 W^{\rm in} x.
\end{eqnarray}
Then each element of the hidden nodes is now processed by 
a nonlinear activation function $\sigma$ such as $\tanh$,
which is denoted by 
\begin{eqnarray} 
\sigma( W^{\rm in} x).
\end{eqnarray}
Finally the output is extracted by an output weight $W^{\rm out}$ ($1 \times N$ dimensional matrix):
\begin{eqnarray} 
 W^{\rm out} \sigma( W^{\rm in} x).
\end{eqnarray}
The parameters in $W^{\rm in}$ and $W^{\rm out}$
are trained such that the error between the output and teacher data becomes minimum.
While this optimization problem is 
highly nonlinear, 
a gradient based optimization, so-called back propagation,
can be employed.
To improve a representation power of the model,
we can concatenate the linear transformation and the activation function as follows:
\begin{eqnarray} 
 W^{\rm out} \sigma \left( W^{(l)} \cdots \sigma \left( W^{(1)} \sigma( W^{\rm in} x) \right) \right),
\end{eqnarray}
which is called multi-layer perceptron or deep neural network.

\subsection{Temporal task}
\label{subsec:temporal}
The above task is not a temporal task, meaning that 
the input data is not sequential but given simultaneously 
like the recognition task of images for 
hand written language, pictures and so on.
However, for a recognition of spoken language or prediction of 
time series like stock market,
which are called temporal tasks, 
the network has to handle the input data that is given in
a sequential way. To do so, the recurrent neural network 
feeds the previous states of the nodes back
into the states of the nodes at next step,
which allows the network to memorize the past input.
In contrast, the neural network without any recurrency
is called a feedforward neural network.

Let us formalize a temporal machine learning task with 
the recurrent neural network.
For given input time series $\{ x_k \}_{k=1}^{L}$
and target time series $\{ \bar y_k \}_{k=1}^{L}$,
a temporal machine learning 
is a task to generalize a nonlinear function,
\begin{eqnarray}
\bar y_k = f(\{ x_j\}_{j=1}^{k}).
\end{eqnarray}
For simplicity,
we consider one-dimensional 
input and output time series,
but their generalization to a multi-dimensional case
is straightforward.
To learn the nonlinear function $f(\{ x_j\}_{j=1}^{k})$,
the recurrent neural network 
can be employed as a model.
Suppose the recurrent neural network 
consists of $m$ nodes and is  
denoted by $m$-dimensional vector 
\begin{eqnarray}
\mb{r}=\left( 
\begin{array}{c}
r_1
\\
\vdots
\\
r_m
\end{array}
\right).
\end{eqnarray}
To process the input time series,
the nodes evolve by
\begin{eqnarray}
\mb{r}(k+1) = \sigma [W \mb{r}(k) +W^{\rm in} x_k],
\end{eqnarray}
where $W$ is an $m \times m$ transition matrix and 
$W^{\rm in}$ is an $m \times 1$ input weight matrix.
Nonlinearity comes from the nonlinear function $\sigma$
applied on each element of the nodes.
The output time series from the network
is defined in terms of a $1 \times m$ readout weights by
\begin{eqnarray}
y_k = W^{\rm out} \mb{r}(k).
\end{eqnarray}
Then the learning task is to determine
the parameters in $W^{\rm in}$, $W$, and $W^{\rm out}$
by using the teacher data $\{ x_k , \bar y_k \}_{k=1}^{L}$
so as to minimize an error between 
the teacher $\{ \bar y_k \}$ and the output $\{ y_k \}$ of the network.

\subsection{Reservoir approach}
While the representation power of the
recurrent neural network can be improved by 
increasing the number of the nodes,
it makes the optimization process of the weights 
hard and unstable. Specifically, the back propagation based 
methods always suffer from the vanishing gradient problem.
The idea of reservoir computing is to resolve 
this problem by mapping an input into a complex higher dimensional 
feature space, i.e., {\it reservoir}, and 
by performing simple linear regression on it.

Let us first see a reservoir approach on a feedforward neural network,
which is called {\it extreme learning machine}~\cite{ELM}.
The input data $x$ is fed into a network like multi-layer perceptron,
where all weights are chosen randomly.
The states of the hidden nodes at some layer is now 
regarded as basis functions of the input $x$ in the feature space:
\begin{eqnarray}
\{ \phi _1 (x), \phi_2 (x),..., \phi_N(x)\}.
\end{eqnarray}
Now the output is defined as a linear combination of these 
\begin{eqnarray}
\sum _i w_i \phi _i (x) + w_0
\end{eqnarray}
and hence the coefficients are determined simply by the linear regression as mentioned before.
If the dimension and nonlinearity of the the basis functions are high enough,
we can model a complex task simply by the linear regression.

The {\it echo state network} is similar but employs the reservoir idea
for the recurrent neural network~\cite{Jaeger0,Maass0,Reservoir},
which has been proposed before extreme learning machine appeared.
To be specific, 
the input weights $W^{\rm in}$ and 
weight matrix $W$ are both chosen randomly  
up to an appropriate normalization.
Then the learning task is done by 
finding the readout weights $W^{\rm out}$
to minimize the mean square error
\begin{eqnarray}
\sum _{k} (y_k - \bar y_k)^2.
\end{eqnarray}
This problem can be solved stably 
by using the pseudo inverse as we mentioned before. 

For both feedforward and recurrent types,
the reservoir approach does not need 
to tune the internal parameters 
of the network depending on the tasks
as long as it posses sufficient complexity.
Therefore, 
the system, to which the machine learning 
tasks are outsourced, is not necessarily 
the neural network anymore, 
but any nonlinear physical system of large degree of freedoms 
can be employed as a {\it reservoir} for information processing,
namely, physical reservoir computing~\cite{Bucket,Laser0,Laser1,Laser1b,Laser2,Laser3,Laser4,Neuromorphic0,Helmut0,Kohei0,Kohei1,Kohei2,Kohei3,Ken1}.

\section{Quantum machine learning on near-term quantum devices}
\label{sec:quantum_machine_learning}
In this section, we will see QRC and related 
frameworks for quantum machine learning.
Before going deep into the temporal tasks done on QRC,
we first explain how complicated quantum natural dynamics
can be exploit as generalization and classification tasks.
This can be viewed as a quantum version of 
extreme learning machine~\cite{QRC_ex}.
While it is an opposite direction to reservoir computing,
we will also see  {\it quantum circuit learning} (QCL)~\cite{QCL},
where the parameters in the complex dynamics is further tuned 
in addition to the linear readout weights.
QCL is a quantum version of a feedforward neural network.
Finally, we will explain quantum reservoir computing 
by extending quantum extreme learning machine for temporal learning tasks.

\subsection{Quantum extreme learning machine}
The idea of quantum extreme learning machine lies in using 
a Hilbert space, where quantum states live, as an enhanced feature space of the input data.
Let us denote the set of input and teacher data by $\{ x^{(j)}, \bar y^{(j)} \}$.
Suppose we have an $n$-qubit system,
which is initialized to 
\begin{eqnarray}
|0\rangle ^{\otimes n}.
\end{eqnarray}
In order to feed the input data into quantum system,
a unitary operation parameterized by $x$, say
$V(x)$,
is applied on the initial state:
\begin{eqnarray}
V(x)|0\rangle ^{\otimes n}.
\end{eqnarray}
For example, if $x$ is one-dimensional data
and normalized to be $0 \leq x \leq 1$,
then we may employ the $Y$-basis rotation $e^{-i \theta Y}$
with an angle $\theta = \arccos (\sqrt{x})$:
\begin{eqnarray}
e^{-i \theta Y} |0\rangle = \sqrt{x} |0\rangle + \sqrt{1-x} |1\rangle.
\end{eqnarray}
The expectation value of $Z$ with respect to $e^{-i \theta Y} |0\rangle$
becomes 
\begin{eqnarray}
\langle  Z \rangle = 2x -1,
\end{eqnarray}
and hence is linearly related to the input $x$.
To enhance the power of quantum enhanced feature space,
the input could be transformed by using a nonlinear function $\phi$:
\begin{eqnarray}
\theta = \arccos(\sqrt{\phi(x)}).
\end{eqnarray}
The nonlinear function $\phi$ could be, for example, 
hyperbolic tangent, Legendre polynomial, and so on.
For simplicity,
below we will use the simple linear input $\theta = \arccos(\sqrt{x})$.

If we apply the same operation on each of the $n$ qubits,
we have 
\begin{eqnarray*}
V(x) |0\rangle ^{\otimes n} &=& (\sqrt{x} |0\rangle + \sqrt{1-x} |1\rangle )^{\otimes n}
\nonumber \\
&=& (1-x)^{n/2} \sum _{i_1,...,i_n}  \prod _k  \sqrt{ \frac{x}{1-x} }^{i_k}  |i_1,...,i_n\rangle .
\end{eqnarray*}
Therefore, we have coefficients that are nonlinear with respect to 
the input $x$ because of the tensor product structure.
Still the expectation value of the single qubit operator $Z_k$ on the $k$th qubit
is $2x-1$. However, if we measure a {\it correlated} operator like $Z_1 Z_2$,
we can obtain a second order nonlinear output
\begin{eqnarray}
\langle Z_1 Z_2 \rangle = (2x-1)^2
\end{eqnarray}
with respect to the input $x$.
To measure a correlated operator, it is enough to apply an entangling unitary operation
like CNOT gate $\Lambda(X)=|0\rangle \langle 0| \otimes I + |1\rangle \langle 1| \otimes X$:
\begin{eqnarray}
\langle \psi | \Lambda_{1,2}(X) Z_1 \Lambda_{1,2}(X) |\psi \rangle  = \langle \psi | Z_1 Z_2 | \psi \rangle.
\end{eqnarray}
In general, an $n$-qubit unitary operation $U$ transforms the observable $Z$
under the conjugation into a linear combination of Pauli operators:
\begin{eqnarray}
U^{\dag} Z_1 U = \sum _{\mb{i}} \alpha _{\mb{i}} P(\mb{i}).
\end{eqnarray}
Thus if you measure the output of the quantum circuit after applying a unitary operation $U$,
\begin{eqnarray}
U V(x) |0\rangle ^{\otimes n},
\end{eqnarray}
you can get a complex nonlinear output, which could be 
represented as a linear combination of exponentially many nonlinear functions.
$U$ should be chosen to be appropriately complex 
with keeping experimental feasibility but not necessarily fine-tuned.

\begin{figure}
\centering
\includegraphics[width=80mm]{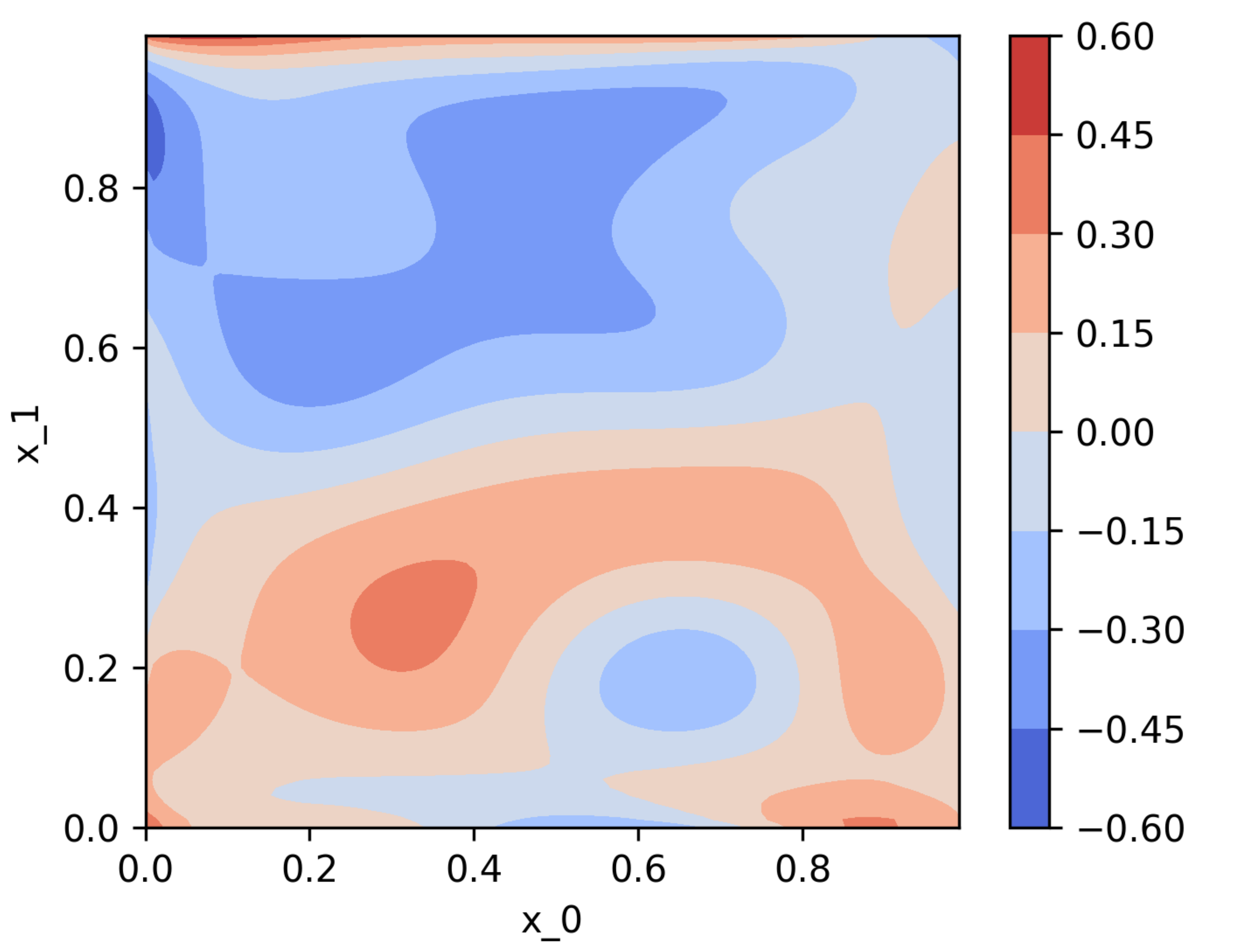}
\caption{The expectation value $\langle Z \rangle$
of the output of a quantum circuit as a function of the input $(x_0,x_1)$.}
\label{fig01}
\end{figure}
To see how the output behaves in a nonlinear way with respect to the input,
in Fig.~\ref{fig01},
we will plot the output $\langle Z \rangle$ for the input $(x_0,x_1)$ and $n=8$,
where the inputs are fed into the quantum state by the $Y$-rotation with angles
\begin{eqnarray}
\theta _{2k} =  k \arccos(\sqrt{x_0})
\\
\theta _{2k+1} =  k \arccos(\sqrt{x_1})
\end{eqnarray}
on the $2k$th and $(2k+1)$th qubits, respectively.
Regarding the unitary operation $U$,
random two-qubit gates are sequentially applied on 
any pairs of two qubits on the 8-qubit system.

\begin{figure}
\centering
\includegraphics[width=80mm]{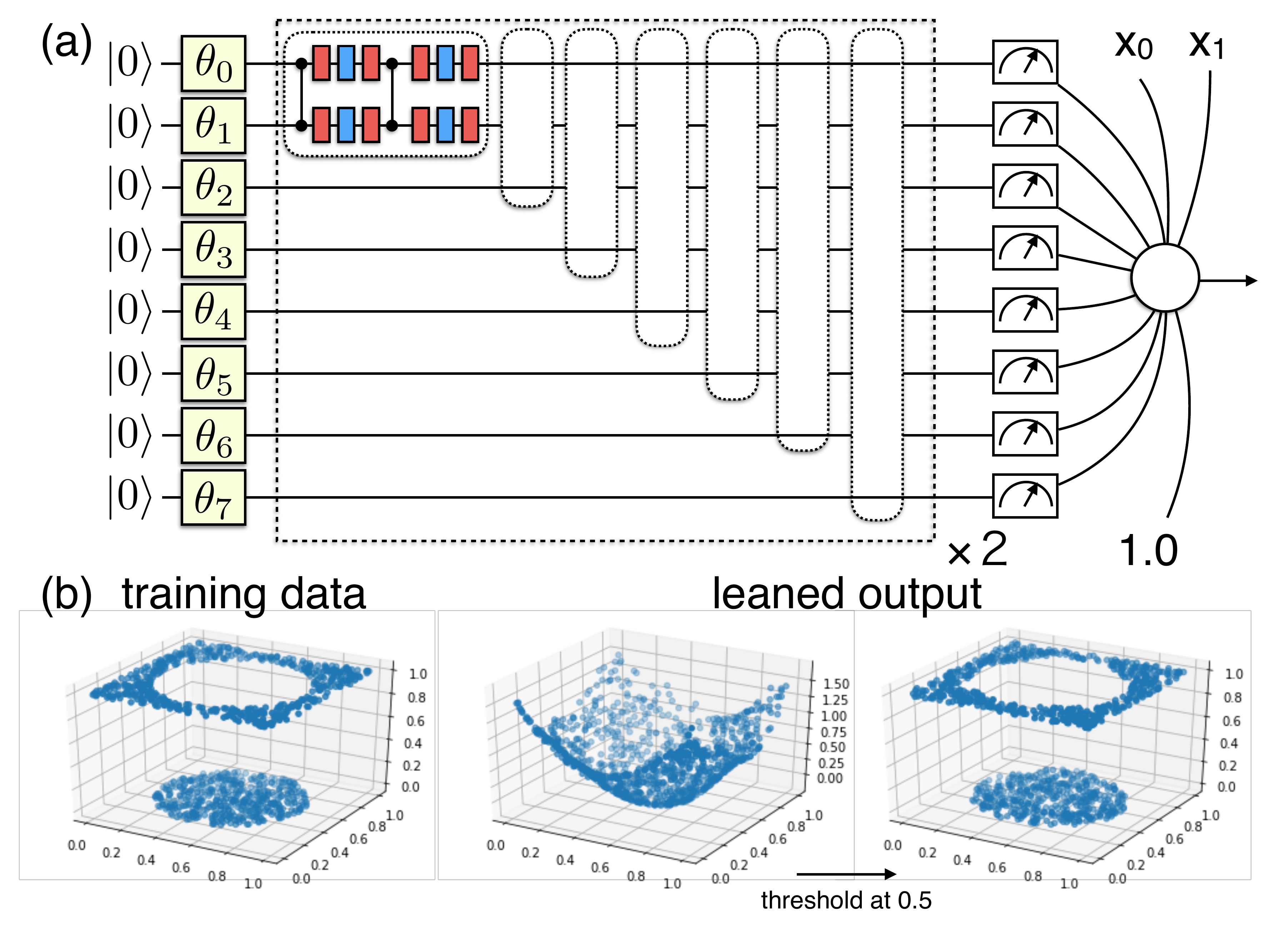}
\caption{(a) The quantum circuit for quantum extreme learning machine. The box with $theta _k$ 
indicates $Y$-rotations by angles $\theta _k$. The red and blue boxes correspond to $X$ and $Z$ rotations by random angles, Each dotted-line box represent a two-qubit gate consisting of two controlled-$Z$ gates and 8 $X$-rotations and 4 $Z$-rotations. As denoted by the dashed-line box, the sequence of the 7 dotted boxes is repeated twice. The readout is defined by a linear combination of $\langle Z_i \rangle$ with 
constant bias term 1.0 and the input $(x_0,x_1)$. (b) (Left) The training data for a two-class classification problem. (Middle) The readout after learning. (Right) Prediction from the readout with threshold at 0.5.}
\label{fig02}
\end{figure}
Suppose the Pauli $Z$ operator is measured on each qubit as an observable.
Then we have 
\begin{eqnarray} 
z_i = \langle Z_i \rangle ,
\end{eqnarray}
for each qubit.
In quantum extreme learning machine, 
the output is defined by taking linear combination of these $n$ output:
\begin{eqnarray}\label{LR}
y = \sum _{i=1}^{n} w_i z_i .
\end{eqnarray}
Now the linear readout weights $\{ w_i \}$ are tuned 
so that the quadratic loss function 
\begin{eqnarray}
L = \sum _j (y^{(j)}- \bar y^{(j)})^2 
\end{eqnarray}
becomes minimum. As we mentioned previously, 
this can be solved by using the pseudo inverse.
In short, quantum extreme learning machine is a linear regression 
on a randomly chosen nonlinear basis functions, 
which come from the quantum state in a space of an exponentially large dimension, 
namely quantum enhanced feature space.
Furthermore, under some typical nonlinear function and unitary operations settings to transform the observables, the output in Eq. (\ref{LR}) can approximate any continuous function of the input. This property is known as the universal approximation property (UAP), which implies that the quantum extreme learning machine can handle a wide class of machine learning tasks with at least the same power as the classical extreme learning machine \cite{UAP_goto}.

Here we should note that a similar approach, quantum kernel estimation, has been taken in Ref.~\cite{IBM_QML_ex}. 
In quantum extreme learning machine, 
a classical feature vector $ \phi_i(x) \equiv \langle \Phi (x) |Z_i | \Phi (x) \rangle$ is extracted from observables on the quantum feature space $|\Phi(x) \rangle \equiv  V(x)|0\rangle ^{\otimes n}$.
Then linear regression is taken by using the classical feature vector.
On the other hand, in quantum kernel estimation, 
quantum feature space is fully employed 
by using support vector machine 
with the kernel functions $K(x,x') \equiv \langle \Phi (x) | \Phi (x') \rangle$, which can be estimated on a quantum computer.
While classification power would be better for quantum kernel estimation,
it requires more quantum computational costs both for learning and prediction in contrast to quantum extreme learning machine.

In Fig.~\ref{fig02},
we demonstrate quantum extreme learning machine for a two-class 
classification task of a two-dimensional input $0\leq x_0, x_1 \leq 1$.
Class 0 and 1 are defined to be those being subject to $(x_0 -0.5)^2 + (x_1-0.5)^2 \leq 0.15$
and $>0.15$, respectively.
The linear readout weights $\{ w_i\}$ are learned with 1000 randomly chosen training data
and prediction is performed with 1000 randomly chosen inputs.
The class 0 and 1 are determined whether or not the output $y$ is larger than 0.5.
Quantum extreme learning machine with an 8-qubit quantum circuit shown in Fig.~\ref{fig02} (a) 
succeeds to predict the class with $95\%$ accuracy.
On the other hand, a simple linear regression for $(x_0,x_1)$ results in $39\%$.
Moreover, quantum extreme learning machine with $U=I$, meaning no entangling gate,
also results in poor, $42\%$. In this way, the feature space enhanced by 
quantum entangling operations is important to obtain a good performance 
in quantum extreme learning machine.

\subsection{Quantum circuit learning}
In the split of reservoir computing, 
dynamics of a physical system is not fine-tuned 
but natural dynamics of the system is harnessed for machine learning tasks.
However, if we see the-state-of-the-art quantum computing devices, 
the parameter of quantum operations can be finely tuned
as done for universal quantum computing.
Therefore it is natural to extend quantum extreme learning machine
by tuning the parameters in the quantum circuit just like 
feedfoward neural networks with back propagation.

Using parameterized quantum circuits for supervised machine leaning tasks 
such as generalization of nonlinear functions and pattern recognitions
have been proposed in Refs.~\cite{QCL,FarhiQNN},
which we call {\it quantum circuit learning}.
Let us consider the same situation with quantum extreme learning machine.
The state before the measurement is given by
\begin{eqnarray}
UV(x) |0\rangle ^{\otimes n}.
\end{eqnarray}
In the case of quantum extreme learning machine
the unitary operation for a nonlinear transformation with respect to the input parameter $x$
is randomly chosen.
However, the unitary operation $U$ may also be 
parameterized:
\begin{eqnarray}
U(\{ \phi_k \}) = \prod _{k} u( \phi_k ).
\end{eqnarray}
Thereby, the output from the quantum circuit
with respect to an observable $A$ 
\begin{eqnarray*}
\langle A (\{ \phi _k\},x) \rangle  = 
\langle 0|^{\otimes n} V^{\dag}(x) U(\{ \phi _k \})^{\dag} 
Z_i U(\{ \phi _k \})V(x) |0 \rangle ^{\otimes n} 
\end{eqnarray*}
becomes a function of the circuit parameters $\{ \phi _k\}$
in addition to the input $x$.
Then the parameters $\{ \phi_k\}$ is tuned 
so as to minimize the error between teacher data and the output,
for example, by using the gradient
just like the output of the feedforward neural network.

Let us define a teacher dataset $\{ x^{(j)}, y^{(j)}\}$
and a quadratic loss function
\begin{eqnarray} 
L(\{ \phi _k\}) = \sum _j (\langle A (\{ \phi _k\},x^{(j)}) \rangle - y^{(j)})^2.
\end{eqnarray}
The gradient of the loss function can be obtained as follows:
\begin{eqnarray*}
\frac{\partial}{\partial \phi _l } L(\{ \phi _k\})
&=&\frac{\partial}{\partial \phi _l } \sum _j (\langle A (\{ \phi _k\},x^{(j)}) \rangle - y^{(j)})^2
\\
&=&\sum _j 2(\langle A (\{ \phi _k\},x^{(j)}) \rangle - y^{(j)}) \frac{\partial}{\partial \phi _l } \langle A (\{ \phi _k\},x^{(j)}) \rangle.
\end{eqnarray*}
Therefore if we can measure the gradient of the observable $\langle A (\{ \phi _k\},x^{(j)}) \rangle$,
the loss function can be minimized according to the gradient descent.

If the unitary operation $u(\phi _k)$ is given by 
\begin{eqnarray}
u(\phi_k) = W_k e^{-i (\phi _k/2) P_k},
\end{eqnarray}
where $W_k$ is an arbitrary unitary, and $P_k$ is a Pauli operator.
Then the partial derivative with respect to the $l$th parameter can be analytically calculated 
from the outputs $\langle A (\{ \phi _k\},x^{(j)}) \rangle$ 
with shifting the $l$th parameter by $\pm \epsilon$  ~\cite{QCL,Mitarai19}:
\begin{eqnarray*}
&&\frac{\partial}{\partial \phi _l } \langle A (\{ \phi _k\},x^{(j)}) \rangle
\nonumber \\
&=& \frac{1}{2 \sin \epsilon} (
\langle A (\{ \phi _1 ,..., \phi _l + \epsilon, \phi _{l+1},... \},x^{(j)}) \rangle \\
&-& \langle A (\{ \phi _1 ,..., \phi _l - \epsilon, \phi _{l+1},... \},x^{(j)}) \rangle).
\nonumber \\
&&
\end{eqnarray*}
By considering the statistical error to measure the observable $\langle A \rangle$,
$\epsilon$ should be chosen to be $\epsilon = \pi/2$ so as to make the denominator maximum.
After measuring the partial derivatives for all parameters $\phi _k$ and calculating 
the gradient of the loss function $L(\{ \phi _k\})$,
the parameters are now updated by the gradient descent:
\begin{eqnarray}
 \theta _l ^{(m+1)} = \theta _l ^{(m)}  -  \alpha \frac{\partial }{\partial \phi _l} L(\{ \phi _k\}).
\end{eqnarray}

The idea of using the parameterized quantum circuits for machine learning  
is now widespread. After the proposal of quantum circuit learning 
based on the analytical gradient estimation above~\cite{QCL} and a similar idea~\cite{FarhiQNN},
several researches have been performed with 
various types of parameterized quantum circuits~\cite{CircuitCentric,TensorNet,Qneural_Chen,generalized_TensorNet,QClassifier_Du} and various models and types of 
machine learning including generative models~\cite{GeneQML_Benedetti,Gene_LiuWang}
and generative adversarial models~\cite{GAN_Situ,GAN_Zeng,GAN_Guzik}.
Moreover, an expression power of the parameterized quantum circuits 
and its advantage against classical probabilistic models have been investigated~\cite{ExpressionPower}.
Experimentally feasible ways to measure an analytical gradient of 
the parameterized quantum circuits have been investigated~\cite{Mitarai19,Xanadu_gradient,CalculusPQC}.
An advantage of using such a gradient for the parameter optimization 
has been also argued in a simple setting~\cite{HarrowGrad},
while the parameter tuning becomes difficult because of the 
vanishing gradient by an exponentially large Hilbert space~\cite{Plateaus}.
Software libraries for optimizing parameterized quantum circuits are now developing
~\cite{Xanadu,VQNet}.
Quantum machine learning on near-term devices,
especially for quantum optical systems,
is proposed in Refs~\cite{Opt_QML,CV_QML}.
Quantum circuit learning with parameterized quantum circuits 
has been already experimentally demonstrated on 
superconducting qubit systems~\cite{IBM_QML_ex,Rigetti_QML_ex} and 
a trapped ion system~\cite{IonQ_QML_ex}.

\subsection{Quantum reservoir computing}
\begin{figure}[t]
\centering
\includegraphics[width=80mm]{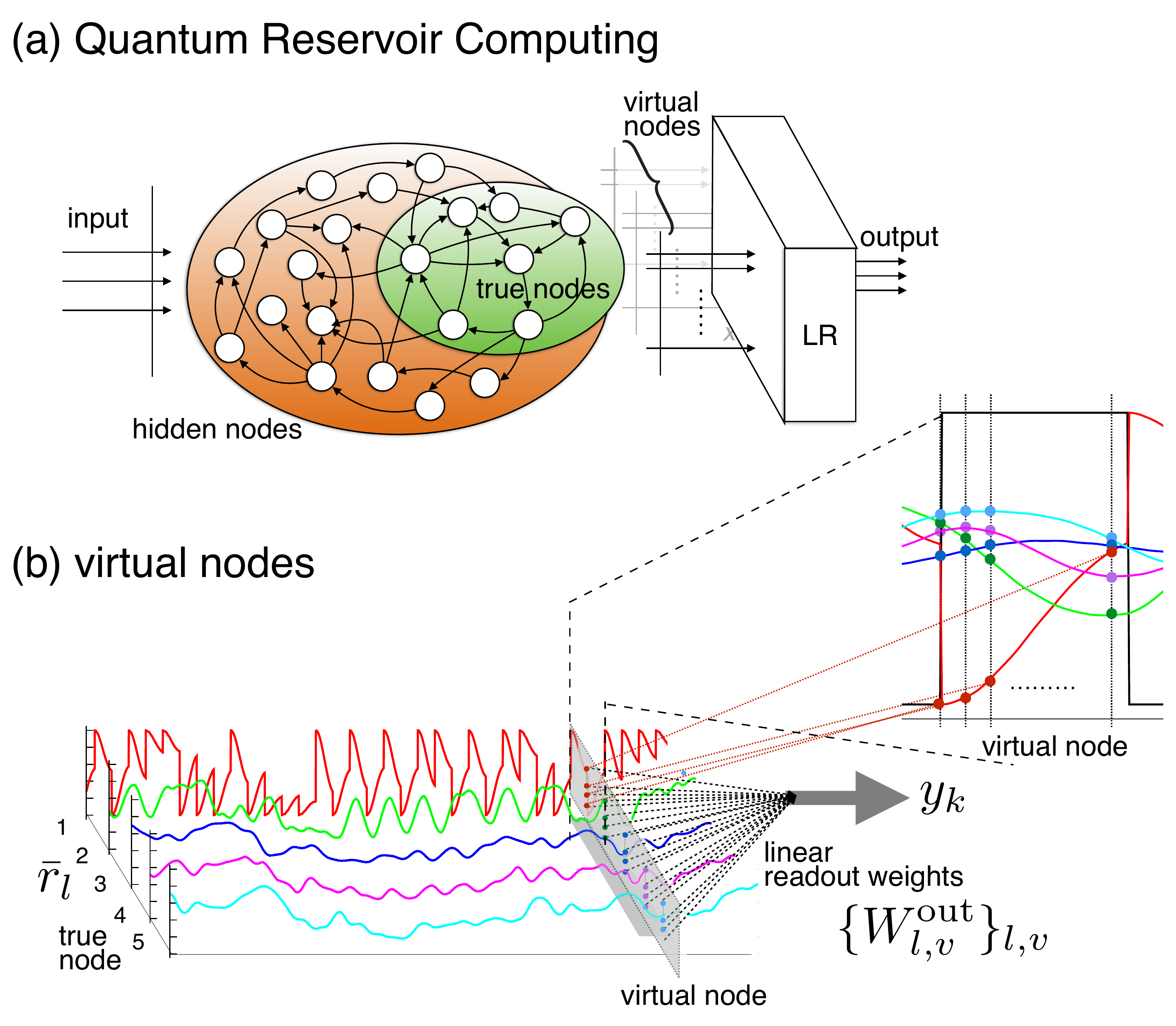}
\caption{(a) Quantum reservoir computing. (b) Virtual nodes and temporal multiplexing.}
\label{fig04}
\end{figure}
Now we return to the reservoir approach and extend 
quantum extreme learning machine from non temporal tasks to temporal ones,
namely, quantum reservoir computing~\cite{QRC}.
We consider a temporal task, which we explained in Sec.~\ref{subsec:temporal}.
The input is given by a time series 
$\{ x_k\}_{k}^{L}$ and the purpose is to learn 
a nonlinear temporal function:
\begin{eqnarray}
\bar y_k = f(\{ x_j \}_{j}^{k}).
\end{eqnarray}
To this end, the target time series $\{ \bar y_k \}_{k=1}^{L}$ 
is also provided as teacher.

Contrast to the previous setting with non temporal tasks,
we have to fed input into a quantum system sequentially.
This requires us to perform an initialization process 
during computation, and hence the quantum state of the system becomes 
mixed state. Therefore, in the formulation of  QRC,
we will use the vector representation of density operators,
which was explained in Sec.~\ref{subsec:vector_density}.

In the vector representation of density operators,
the quantum state of an $N$-qubit system is given by 
a vector in a $4^N$-dimensional real vector space,
$\mb{r} \in \mathbb{R}^{4^N}$.
In QRC, similarly to recurrent neural networks, 
each element of the $4^N$-dimensional vector is
regarded as a {\it hidden} node of the network.
As we seen in Sec.~\ref{subsec:vector_density}, 
any physical operation can be written as a 
linear transformation of the real vector by a $4^N \times 4^N$ matrix $W$:
\begin{eqnarray}
\mb{r}' = W \mb{r}.
\label{eq:time_evolution}
\end{eqnarray}

Now we see, from Eq.~(\ref{eq:time_evolution}), a time evolution
similar to the recurrent neural network,
$\mb{r}' = \tanh( W \mb{r} )$.
However,
there is no nonlinearity such as $\tanh$ in 
each quantum operation $W$.
Instead, 
the time evolution $W$
can be changed according to 
the external input $x_k$,
namely $W_{x_k}$,
which contrasts to the conventional 
recurrent neural network 
where the input is fed additively
$W \mb{r} + W^{\rm in} x_k$.
This allows the quantum reservoir 
to process the input information $\{x_k\}$
nonlinearly, by repetitively 
feeding the input.

Suppose the input $\{ x_k\}$ is 
normalized such that $0 \leq x_k \leq 1$.
As an input,
we replace a part of the qubits
to the quantum state.
The density operator is given by 
\begin{eqnarray}
\rho _{x_k} = \frac{I+(2 x_k - 1)Z}{2}.
\end{eqnarray}
For simplicity,
below we consider the case where 
only one qubit is replaced for the input.
Corresponding matrix $S_{x_k}$
is given by 
\begin{eqnarray*}
(S_{x_k})_{\mb{j}\mb{i}} = {\rm Tr}\left\{ P(\mb{j}) \frac{I+(2x_k-1)Z}{2}  
\otimes {\rm Tr}_{\rm replace} [P(\mb{i})] \right\} /2^N,
\end{eqnarray*}
where ${\rm Tr}_{\rm replace}$ indicates 
a partial trace with respect to the replaced qubit.
With this definition, we have
\begin{eqnarray}
\rho ' =  {\rm Tr}_{\rm replace}[\rho] \otimes \rho _{x_k}  
\Leftrightarrow \mb{r}'  = S_{x_k} \mb{r}.
\end{eqnarray}

The unitary time evolution,
which is necessary to obtain a nonlinear behavior 
with respect to the input valuable $x_k$,
is taken as a Hamiltonian dynamics
$e^{-i H \tau}$ for a given time interval $\tau$.
Let us denote its representation on the vector space by $U_\tau$:
\begin{eqnarray}
\rho' = e^{-i H \tau} \rho e^{i H \tau}
\Leftrightarrow \mb{r}' = U_{\tau} \mb{r}.
\end{eqnarray}
Then, a unit time step is written as
an input-depending linear transformation:
\begin{eqnarray}
\mb{r}((k+1)\tau) =  U_\tau S_{x_k}  \mb{r}(k\tau).
\end{eqnarray}
where $\mb{r}(k\tau)$ indicates 
the hidden nodes at time $k\tau$.

Since the number of the hidden nodes 
are exponentially large,
it is not feasible to observe all nodes 
from experiments.
Instead, a set of observed nodes $\{\bar r_l\}_{l=1}^{M}$,
which we call {\it true nodes}, is 
defined by a $M \times 4^N $ matrix $R$,
\begin{eqnarray}
\bar r_l (k\tau) = \sum _{\mb{i}} R_{l\mb{i}}  r_{\mb{i}}(k\tau).
\end{eqnarray}
The number of true nodes $M$ has to be a polynomial in the number of qubits $N$.
That is, 
from exponentially many hidden nodes,
a polynomial number of true nodes are obtained 
to define the output from QR (see Fig.~\ref{fig04} (a)):
\begin{eqnarray}
y_k = \sum _l W^{\rm out}_l \bar r_l (k\tau),
\end{eqnarray}
where $W_{\rm out}$ is the readout weights,
which is obtained by using the training data.
For simplicity, 
we take the single-qubit Pauli $Z$ operator
on each qubit as the true nodes, i.e.,
\begin{eqnarray}
\bar r_l = {\rm Tr}[Z_l \rho],
\end{eqnarray}
so that if there is no dynamics these nodes 
simply provide a linear output $(2x_k -1)$ with respect to the input $x_k$.

Moreover, 
in order to improve the performance 
we also perform the temporal multiplexing.
The temporal multiplexing has been found to be 
useful to extract complex dynamics on the exponentially 
large hidden nodes
through the restricted number of the true nodes~\cite{QRC}.
In temporal multiplexing,
not only the true nodes just 
after the time evolution $U_\tau$,
also at each of the subdivided $V$ time intervals during the unitary evolution 
$U_{\tau}$ to construct $V$ virtual nodes, as shown in Fig.~\ref{fig04} (b). 
After each input by $S_{x_k}$, 
the signals from the hidden nodes (via the true nodes)
are measured for each subdevided intervals
after the time evolution by $U_{ v \tau /V}$
($v=1,2,...V$), i.e., 
\begin{eqnarray}
\mb{r}(k\tau +(v/V)\tau) \equiv U_{(v/V)\tau} S_{x_{k}} \mb{r}(k\tau).
\end{eqnarray}
In total, now we have $N \times V$ nodes,
and the output is defined as their linear combination:
\begin{eqnarray}
y_k = \sum _{l=1}^{N} \sum _{v=1}^{V} W^{\rm out}_{j,v} \bar r_{l}(k\tau +(v/V)\tau).
\end{eqnarray}
By using the teacher data $\{ \bar y_k \}_{k}^{L}$,
the linear readout weights $W^{\rm out}_{j,v}$ can be determined 
by using the pseudo inverse.
In Ref.~\cite{QRC},
the performance of QRC has been investigated extensively
for both binary and continuous inputs.
The result shows that even if the number of the qubits are small
like 5-7 qubits the performance as powerful as the echo state network of the 
100-500 nodes have been reported both in short term memory and 
parity check capacities.
Note that, although we do not go into detail in this chapter, the technique called spatial multiplexing~\cite{QRC_multiplex}, which exploits multiple quantum reservoirs with common input sequence injected, is also introduced to harness quantum dynamics as a computational resource.
Recently, QRC has been further investigated in Refs.~\cite{QRC_OTOC,QRC_sig,QRC_aus}.
Specifically, in Ref.~\cite{QRC_sig}, 
the authors use quantum reserovir computing to 
detect many-body entanglement by estimating nonlinear functions
of deinsity operators like entropy.

\subsection{Emulating chaotic attractors using quantum dynamics}
\begin{figure*}
\centering
\includegraphics[width=140mm]{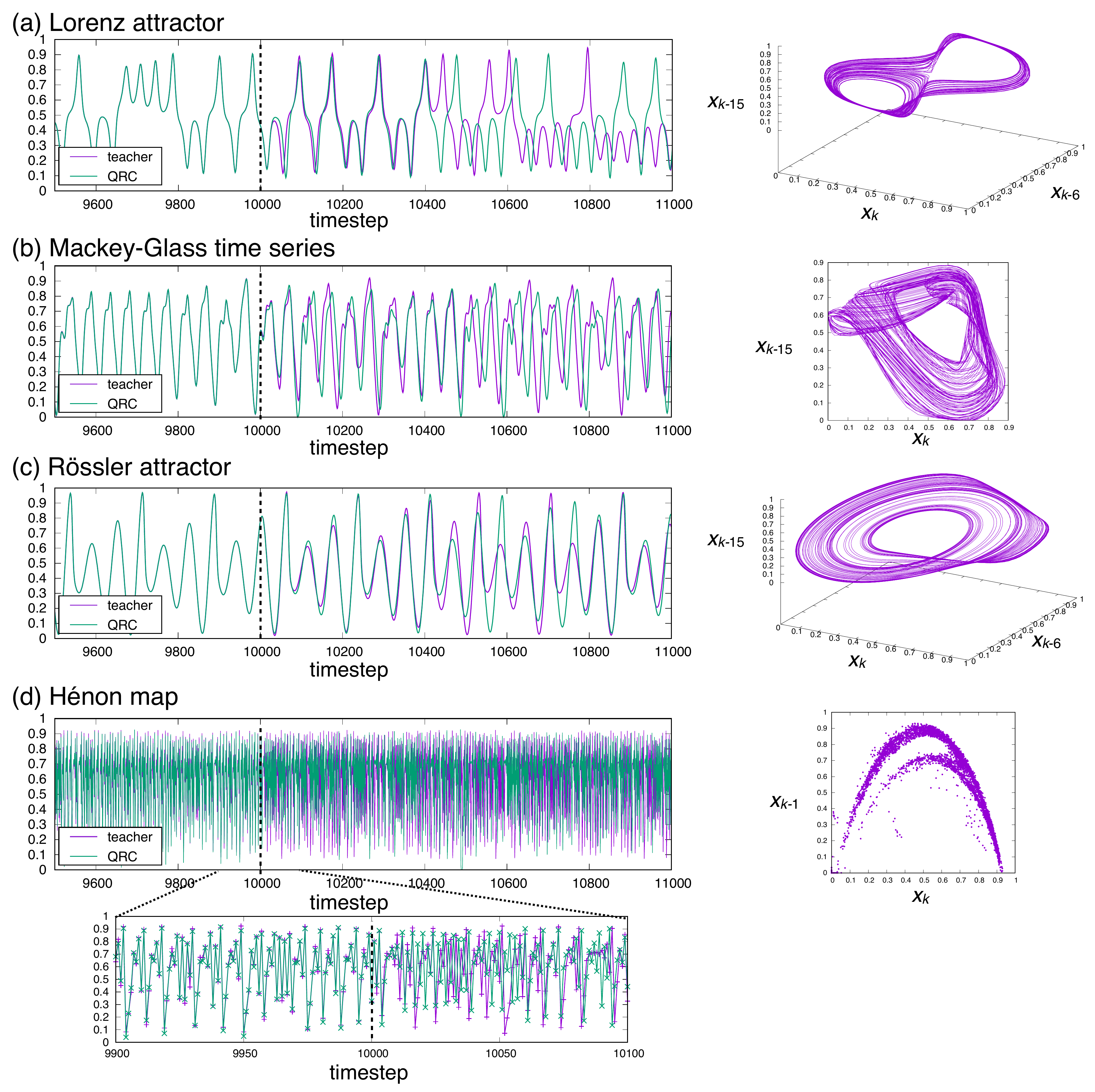}
\caption{Demonstrations of chaotic attractor emulations. (a) Lorenz attractor. (b) Mackey-Glass system. (c) R{\"o}ssler attractor. (d) H{\'e}non map. The dotted line shows the time step when the system is switched from teacher forced state to autonomous state. In the right side, delayed phase diagrams of learned dynamics are shown.}
\label{fig03}
\end{figure*}
To see a performance of QRC,
here we demonstrate an emulation of chaotic attractors.
Suppose $\{ x_k\}_k^{L}$ is a discretized time sequence being 
subject to a complex nonlinear equation, which might has a chaotic behavior.
In this task,
the target, which the network is to output, is 
defined to be 
\begin{eqnarray}
\bar y_{k} = x_{k+1} = f(\{x_{j}\}_{j=1}^{k}).
\end{eqnarray}
That is, the system learns the input of the next step.
Once the system successfully learns $\bar y_{k}$,
by feeding the output into the input of the next step of the system,
the system evolves autonomously.

Here we employ the following
target time series from chaotic attractors:
(i) Lorenz attractor, 
\begin{eqnarray}
\frac{dx}{dt} &=& a (y-x),
\\
\frac{dy}{dt} &=& x (b-z) - y,
\\
\frac{dz}{dt} &=& xy - cz,
\end{eqnarray}
with $(a,b,c)=(10,28,8/3)$,
(ii) the chaotic attractor of Mackey-Glass equation, 
\begin{eqnarray}
\frac{d}{dt} x(t) = \beta \frac{x(t-\tau)}{1+x(t-\tau)^n} - \gamma x(t)
\end{eqnarray}
with $(\beta , \gamma , n) = (0.2, 0.1 , 10)$ and $\tau = 17$,
(iii) R{\"o}ssler attoractor, 
\begin{eqnarray}
\frac{dx}{dt} &=& -y-z,
\\
\frac{dy}{dt} &=& x+ay,
\\
\frac{dz}{dt} &=& b+z(x-c),
\end{eqnarray}
with $(0.2,0.2,5.7)$, 
and (iv)
H{\'e}non map, 
\begin{eqnarray}
x_{t+1} = 1-1.4x_t +0.3x_{t-1}.
\end{eqnarray}
Regarding (i)-(iii),
the time series is obtained by using the fourth-order Runge-Kutta method
with step size 0.02,
and only $x(t)$ is employed as a target.
For the time evolution of quantum reservoir,
we employ a fully connected transverse-field Ising model
\begin{eqnarray} 
H = \sum _{ij} J_{ij} X_i X_j + h Z_i,
\label{eq_Ising}
\end{eqnarray}
where the coupling strengths are randomly chosen such that $J_{ij}$ 
is distributed randomly from $[-0.5,0.5]$ and $h=1.0$.
The time interval and the number of the virtual nodes are chosen to be $\tau = 4.0$ and $v=10$ 
so as to obtain the best performance.
The first $10^4$ steps are used for training. After the linear readout weights 
are determined, several $10^3$ steps are predicted by autonomously 
evolving the quantum reservoir.
The results are shown in Fig.~\ref{fig03} for each of (a) Lorenz attractor,
(b) the chaotic attractor of Mackey-Glass system, (c) R{\"o}ssler attractor, and (d) H{\'e}non map.
All these results show that training is done well and the prediction
is successful for several hundreds steps.
Moreover, the output from the quantum reservoir also successfully 
reconstruct the structures of these chaotic attractors
as you can see from the delayed phase diagram.

\section{Conclusion and Discussion}
\label{sec:conclusion}
Here we reviewed quantum reservoir computing 
and related approaches, quantum extreme learning machine and 
quantum circuit learning.
The idea of quantum reservoir computing comes from 
the spirit of reservoir computing, i.e., 
outsourcing information processing to natural physical systems.
This idea is best suited to quantum machine learning 
on near-term quantum devices in NISQ (noisy intermediate quantum) era.
Since reservoir computing uses complex physical systems
as a feature space to construct a model by the simple linear regression,
this approach would be a good way to understand
the power of a quantum enhanced feature space.

\section*{Acknowledgement}
KF is supported by KAKENHI No.16H02211, JST PRESTO JPMJPR1668, JST ERATO JPMJER1601, and JST CREST JPMJCR1673. 
KN is supported by JST PRESTO Grant Number JPMJPR15E7, Japan, by JSPS KAKENHI Grant Numbers JP18H05472, JP16KT0019, and JP15K16076. 
KN would like to acknowledge Dr. Quoc Hoan Tran for his helpful comments.
This work is supported by MEXT Quantum Leap Flagship Program (MEXT Q-LEAP) Grant No. JPMXS0118067394.

\end{document}